\documentclass[journal]{IEEEtran}

\usepackage{amsmath,amsfonts}
\usepackage{algorithmic}
\usepackage{algorithm}
\usepackage{array}
\usepackage{textcomp}
\usepackage{stfloats}
\usepackage{multicol}
\usepackage{url}
\usepackage{siunitx}
\usepackage{mathtools}
\usepackage{verbatim}
\usepackage{graphicx}
\usepackage{cite}
\usepackage{amsmath}
\usepackage{subcaption}

\DeclareMathOperator{\grd}{grad}
\DeclareMathOperator{\crl}{curl}
\DeclareMathOperator{\dvg}{div}

\DeclareMathOperator{\RT}{RT}
\DeclareMathOperator{\BC}{BC}
\DeclareMathOperator*{\spn}{span}

\newcommand{\bbRT}{\mathbb{RT}}
\newcommand{\bbBC}{\mathbb{BC}}

\newcommand{\vct}[1]{\boldsymbol{\mathsf{#1}}}
\newcommand{\mat}[1]{\boldsymbol{\mathsf{#1}}}

\title{Fast Converging Single Trace Quasi-local PMCHWT Equation for the Modelling of Composite Systems}

\author{Kristof Cools, \IEEEmembership{Member, IEEE}
\thanks{This work was supported by the European Research Council (ERC) under the European Union’s Horizon 2020 research and innovation programme (Grant agreement No. 101001847).}
\thanks{Kristof Cools is with the dept. of information technology, Ghent University, Ghent, 9052 BE (e-mail: kristof.cools@ugent.be).}
}

\IEEEpubid{}

\begin{document}

\maketitle

\begin{abstract}
The PMCHWT integral equation enables the modelling of scattering of time-harmonic fields by penetrable, piecewise homogeneous, systems. They have been generalised to include the modelling of composite systems that may contain junctions, i.e. lines along which three or more materials meet. Linear systems resulting upon discretisation of the PMCHWT are, because of their large dimension, typically solved by Krylov iterative methods. The number of iterations required for this solution critically depends on the eigenvalue distribution of the system matrix. For systems that do not contain junction lines, Calder\'on preconditioning, which was first applied to the electric field integral equation, has been generalised to the PMCHWT equation. When junctions are present, this approach cannot be applied. Alternative approaches, such as the global multi-trace method, conceptually remove the junction lines and as a result are amenable to Calder\'on preconditioning. This approach entails a doubling of the degrees of freedom, and the solution that is produced only approximately fulfils the continuity conditions at interfaces separating domains. In this contribution, a single trace quasi-local PMCHWT equation is introduced that requires a number of iterations for its solution that only slowly increases as the mesh size tends to zero. The method is constructed as a generalisation of the classic PMCHWT, and its discretisation is thoroughly discussed. A comprehensive suite of numerical experiments demonstrates the correctness, convergence behaviour, and efficiency of the method. The integral equation is demonstrated to be free from interior resonances.

\end{abstract}


\section{Introduction}

The modelling of time-harmonic electromagnetic fields in penetrable systems built by arranging homogenous regions in general geometric configurations poses a number of challenges compared to simpler scenarios such as the scattering of waves by a single perfectly conducting surface. It requires the computation of the tangential components of both the electric and magnetic field at the boundary of all participating domains, and the enforcement of field continuity at the interfaces between neighbouring domains, either as an essential or as a natural condition.

Modellers have developed boundary element methods that are based on trial and test finite element spaces spanned by bases of locally supported functions that are designed to all exhibit the required continuity conditions \cite{yla-oijala_surface_2005, kolundzija_electromagnetic_1999}. These methods are optimally accurate and have been generalised to be able to accommodate all combinations of penetrable media and perfectly conducting regions that are encountered in practice. Unfortunately, the resulting linear systems have a highly complicated eigenvalue spectrum, especially in the presence of junctions (lines  along which three or more domains meet) and when highly accurate results are required. These systems cannot be solved by iterative solvers of Krylov type (and thus cannot benefit from matrix-vector compression methods such as the fast multipole method and H-matrix methods). Reverting to LU to solve the system results in a computational complexity that is cubic in the number of unknowns.

To get a grip on the spectrum of the BEM system and thus enable its efficient solution by iterative methods, the transmission problem has been cast in equivalent forms that are more suitable for this pursuit. Most relevant to this work are the multi-trace methods. The global multi-trace method is based on the introduction of a conceptual gap between all bounded domains. This gap can be thought of as being occupied by the background material. This process essentially eliminates junctions and leads to integral equations that can be stated in terms of the action of integral operators on currents supported by the complete boundaries of the participating domains. This implies that low frequency stabilisation methods such as loop-tree splitting and dense grid stabilisation such as Calder\'on preconditioning now become available. Global multi-trace formulations have been explored in \cite{yunhui_chu_surface_2003} for the low-frequency stabilisation of the boundary element method and for the dense grid stabilisation in \cite{yla-oijala_calderon_2011}. In \cite{claeys_electromagnetic_2012} well-posedness has been established, and in \cite{lasisi_fast_2022} dense grid stabilisation through Calder\'on preconditioning was extended to allow inclusion of screens and metallic bodies. Dense grid stabilisation using multi-resolution bases is explored in \cite{martin_multitrace_2023}.

Another type of formulations are the local multi-trace formulations. The formulations enforce a judiciously chosen linear combination of the extinction property in each participating domain with a criterion for the continuity of fields (dubbed the transmission condition) on the interface of the domain under consideration with each of its neighbours. The resulting system is typically solved by a block Jacobi or block Gauss-Seidel method \cite{peng_computations_2013}. The efficiency of this approach and in particular the convergence rate of the iterative process depends critically on the choice of discretisation and transmission condition \cite{peng_novel_2015}.

Global and local multi-trace methods are effective and have enabled the modelling of highly complicated systems. In some cases they are not the preferred method because they introduce twice the number of unknowns compared to single trace methods or simply because solutions are required that exactly meet the continuity conditions at the interfaces between domains. Especially in multi-physics scenarios or when coupling to legacy software, data computed in one domain needs to be transferred to neighbouring domains without loss of fidelity.

The application domain of Calder\'on preconditioners for single-trace methods has been expanded in the last decades to include many particle systems \cite{kleanthous_calderon_2019}, layered media \cite{chen_calderon_2014}, and periodic media \cite{niino_preconditioning_2012}. Other contributions focused on their further optimisation and implementation aspects \cite{niino_calderon_2012, yan_comparative_2010}. For single domain problems regularisers aimed at low-frequency problems have been developed, both for lossless \cite{beghein_low-frequency_2017} and lossy materials \cite{giunzioni_low-frequency_2024}. For the second kind M\"uller equation (which is inherently well-conditioned), single trace equations exist that lead to rapidly converging methods, even when junctions are present \cite{claeys_second-kind_2017}, set in an $L^2$ functional framework. The dense grid stabilisation of first kind single-trace methods that is effective in the presence of junctions has remained an open problem.

This paper introduces the \emph{quasi-local PMCHWT} (QL-PMCHWT), a single-trace method for the modelling of the transmission of time-harmonic electromagnetic fields in systems that are composed from piecewise homogeneous penetrable materials. The QL-PMCHWT produces linear systems whose solution by Krylov iterative solvers requires a number of iterations that grows only slowly as the mesh size $h$ tends to zero. This is important, especially when multi-scale geometries are present or when highly accurate solutions are required. One the oter hand, the method produces approximate solutions that converge in energy at least linearly in $h$ to the exact solution. The method allows only a single unique solution for all real frequencies $\omega$, i.e. it is not vulnerable to the problem of interior resonances. By construction, the solutions meet the continuity conditions at the interfaces exactly.

The paper starts with a revision of the classic PMCHWT for composite systems containing junctions, setting language and notation that facilitates the further exposition. Next, the quasi-local PMCHWT is introduced as a generalisation of the classic PMCHWT, using a different test space and a parametrisation of this test space that is chosen to achieve the regularisation in the dense grid regime. Finally, a comprehensive set of examples is provided, aimed at verifying the correctness, efficiency, and immunity to resonances of the quasi-local PMCHWT.

\section{The classic single trace PMCHWT}
\label{sec:classic_PMCHWT}

Consider the partition of space by disjoint open sets $\mathbb{R}^3 = \cup_{i=0}^N \bar{\Omega}_i$, where only $\Omega_0$ is unbounded. The boundaries $\partial \Omega_i$ are imbued with an outward pointing unit normal $n_i$. The skeleton $\cup_{i=0}^N \partial \Omega_i$ is denoted $\Gamma$ and the interfaces $\Gamma_ij$ are open sets such that $\bar{\Gamma}_{ij} = \partial \Omega_i \cap \partial \Omega_j$.

All fields are considered phasors $F$ that vary harmonic in time as $\Re(F e^{i\omega t})$ with angular frequency $\omega$. Domain $\Omega_i$ is occupied by a material characterised by permittivity $\epsilon_i$ and permeability $\mu_i$, or equivalently wavenumber $\kappa_i=\omega \sqrt{\epsilon_i \mu_i}$ and impedance $\eta_i = \sqrt{\mu_i / \epsilon_i}$. Excitations $e^{inc}_i, h^{inc}_i$ with $i=0,...,N$ are solutions to the time-harmonic Maxwell equations w.r.t. material $(\epsilon_i, \mu_i)$ with sources supported within $\Omega_i$ (plane waves are interpreted as being generated by sources at infinity). The solution to the transmission problem is the total field $(e_i, h_i)_{i=0}^N := \left( \left. e \right|_{\Omega_i}, \left .h \right|_{\Omega_i} \right)_{i=0}^N$ such that:
\begin{itemize}
    \item The scattered fields $(e_i-e^{inc}_i, h_i-h^{inc}_i)$ are solutions to the sourceless Maxwell equations with respect to the material occupying $\Omega_i$, $i=0,...,N$.
    \item The tangential components $(e_i \times n_i, n_i \times h_i)$ are continuous at the interfaces $\Gamma_{ij} := \bar{\Omega}_i \cap \bar{\Omega}_j$.
    \item The scattered field $(e^{sc}_0, h^{sc}_0) := (e_0 - e^{inc}_0, h_0 - h^{inc}_0)$ fulfils the Silver-M\"uller radiation condition.
\end{itemize}

The PMCHWT applied to geometries that include junctions can best be understood in the framework of multi-trace spaces \cite{claeys_electromagnetic_2012}. The multi-trace function space that provides a starting point for this exploration is $\mathbb{H}^{-1/2}(\dvg, \Gamma) := \prod_{i=0}^N {\mathbf{H}}^{-1/2}(\dvg, \partial \Omega_i)$ with $\mathbf{H}^{-1/2}(\dvg, \partial \Omega_i) = H^{-1/2}(\dvg, \partial \Omega_i) \times H^{-1/2}(\dvg, \partial \Omega_i)$ (including a factor for both the electric and magnetic traces) \cite{claeys_electromagnetic_2012}. The single trace space is denoted $\mathbf{H}^{-1/2}(\dvg, \Gamma)$ and is the subspace of this multi-trace space that corresponds to pairs of electric and magnetic fields whose tangential components are continuous at all interfaces. In this paper, the natural isomorphism between $\prod_{i=0}^N {\mathbf{H}}^{-1/2}(\dvg, \partial \Omega_i)$ and $\prod_{i=0}^N {H}^{-1/2}(\dvg, \partial \Omega_i) \times \prod_{i=0}^N {H}^{-1/2}(\dvg, \partial \Omega_i)$ will be tacitly relied upon, i.e. trace field components are grouped according to their domain or kind (electric/magnetic), whichever is more convenient.

In any of the domains $\Omega_i$, $i=0,...N$, if the pair of traces $(m_i, j_i) :=(e_i \times n_i, n_i \times h_i)$ belongs to a solution in that domain with respect to its assigned material, it holds that
\begin{multline}
    \label{eq:ext_ops}
    A_i (m_i, j_i) := \\
    \left( \begin{array}{cc}
        -\frac{1}{2} - K_i & \eta_i T_i \\
        -\eta_i^{-1} T_i & -\frac{1}{2} - K_i
    \end{array} \right) \left( \begin{array}{c}
        m_i \\
        j_i
    \end{array} \right) = -\left( \begin{array}{c}
        e^{inc}_i \times n_i \\
        n_i \times h^{inc}_i
    \end{array} \right),
\end{multline}
where $T_i$ and $K_i$ are the single and double layer boundary integral operators defined by
\begin{align}
    T_i (m)(x) = & -\iota \kappa_i n_i \times \int_\Gamma \frac{e^{-\iota \kappa_i |x-y|}}{4\pi |x-y|} m(y) dy + \notag \\
    = & \frac{1}{\iota \kappa_i} n_i \times \grd \int_\Gamma \frac{e^{-\iota \kappa_i |x-y|}}{4\pi |x-y|} \dvg m(y) dy \notag, \\
    K_i (m)(x) = & n_i \times \operatorname{pv} \int_\Gamma \grd_x \frac{e^{-\iota \kappa_i |x-y|}}{4\pi |x-y|} \times m(y)dy. \notag
\end{align}
Here, $\operatorname{pv}$ stands for the Cauchy principal value, $\grd$ is the spatial gradient, and $\dvg$ is the surface divergence. For a tuple of traces $(m,j) := (m_i, j_i)_{i=0}^N$ in which all entries fulfil \eqref{eq:ext_ops}, it holds for all $(p,q) = (p_i,q_i)_{i=0}^N$ in $\mathbb{H}^{-1/2}(\dvg, \Gamma)$ that
\begin{multline}
    \label{eq:ext_var}
    \sum_{i=0}^N \left< (p_i,q_i), (-1/2-A_i)(m_i,j_i) \right>_\times = \\
    \sum_{i=0}^N \left< (p_i,q_i), (e^{inc}_i \times n_i), n_i \times h^{inc}_i \right>_\times
\end{multline}
with $\left< (p_i,q_i), (r_i,s_i) \right>_\times := \int_{\partial \Omega_i} (n_i \times p_i) \cdot r_i + \int_{\partial \Omega_i} (n_i \times q_i) \cdot s_i$ and $\left< (p,q), (r,s) \right>_\times := \sum_{i=0}^N \left< (p_i,q_i), (r_i,s_i) \right>_\times$. The \emph{block-diagonal} operator defined by the left hand size of \eqref{eq:ext_var} is denoted $-1/2-A$.

It is important to note that if both $(p,q)$ and $(m,j)$ are in the single trace space, the so-called self-polarity property holds:
\begin{equation}
    \left< (p,q), (j,m) \right>_\times = 0.
\end{equation}
For the single domain case, this property is equivalent to noticing that when subtracting the inner and outer Calder\'on projectors the identity operators in \eqref{eq:ext_ops} cancel. It gives an exact meaning to the notion of subtracting functions that are defined on logically different domains as well as a generalisation to the multi-domain case and in particular when junctions are present. The basic intuition remains the same: the single trace functions $(p,q)$ and $(j,m)$ both change sign upon crossing the interfaces $\Gamma_{ij}$; the additional sign flip of the normals results in the cancelation of contributions from opposite sides.

The self-polarity property implies that in the single trace PMCHWT, the identity contributions in $\eqref{eq:ext_var}$ can be left out, simplifying both the analysis of the method and its implementation.

This leads to the formulation of the continuous single trace PMCHWT equation: find $(m,j) \in \mathbf{H}^{-1/2}(\dvg, \Gamma$), such that for all $(p,q) \in \mathbf{H}^{-1/2}(\dvg, \Gamma)$ the equality \eqref{eq:ext_var} holds.

The discretisation of the single trace PMCHWT starts by defining a triangular mesh $\mathcal{T}_h$ on the skeleton that is geometrically conforming so that by restriction the global mesh induces meshes $\mathcal{T}_{h,i}$ for the domain boundaries $\partial \Omega_i$. When required, the orientation of the triangles making up $\mathcal{T}_{h,i}$ are changed to reflect the orientation of $\partial \Omega_i$.

In what follows, the choice of basis, its ordering, and the sign convention will be important. For each of the meshes $\mathcal{T}_h$ and $\mathcal{T}_{h,i}$, denote a particular ordering and orientation of the set of edges by $\mathcal{E}_h$ and $\mathcal{E}_{h,i}$, respectively.

The space $\RT(\mathcal{T}_{h,i})$ is spanned by the ordered basis $\left( f^{i}_{e} \right)_{e \in \mathcal{E}_{h,i}}$ of RWG functions \cite{rao_electromagnetic_1982} such that $\int_{e^\prime} \left( f^i_e \times n_i \right) \cdot t_{e^\prime} = \delta_{e^\prime e}$ with $t_{e^\prime}$ the unit tangent to the oriented edge $e$ and $\delta_{e^\prime e}$ the Kr\"onecker delta. Note that even though the meshes $\mathcal{T}_{h,i}$ are built by restricting the skeleton mesh $\mathcal{T}_h$ to $\partial \Omega_i$, there is in general no obvious relation between their respective sets of edges.

A finite element space for the single trace space is constructed as a subspace of the finite element space for the multi-trace space
\begin{equation}
    \bbRT(\mathcal{T}_h) := \prod_{i=0}^{N} \mathbf{RT}(\mathcal{T}_{h,i})
\end{equation}   
where $\mathbf{RT}(\mathcal{T}_{h,i}) = \RT
(\mathcal{T}_{h,i}) \times \RT
(\mathcal{T}_{h,i})$. These direct product spaces inherit their bases (which from here on will be referred to as the \emph{standard basis}) from their \emph{atomic} constituents $\RT(\mathcal{T}_{h,i})$. For the construction of the regulariser in section~\ref{sec:qlpmchwt}, a dual to this finite element space is required. Starting from the space $\BC(\mathcal{T}_{h,i})$ of Buffa-Christiansen functions on $\mathcal{T}_{h,i}$ (with a basis informed by the ordering and orientation of edges in $\mathcal{E}_{h,i}$), the construction is completely analogue, leading to the definition $\bbBC(\mathcal{T}_h) := \prod_{i=0}^N \mathbf{BC}(\mathcal{T}_{h,i})$, with  $\mathbf{BC}(\mathcal{T}_{h,i}) := \BC(\mathcal{T}_{h,i}) \times \BC(\mathcal{T}_{h,i})$.

\begin{figure}
    \centering
    \includegraphics[width=1\linewidth]{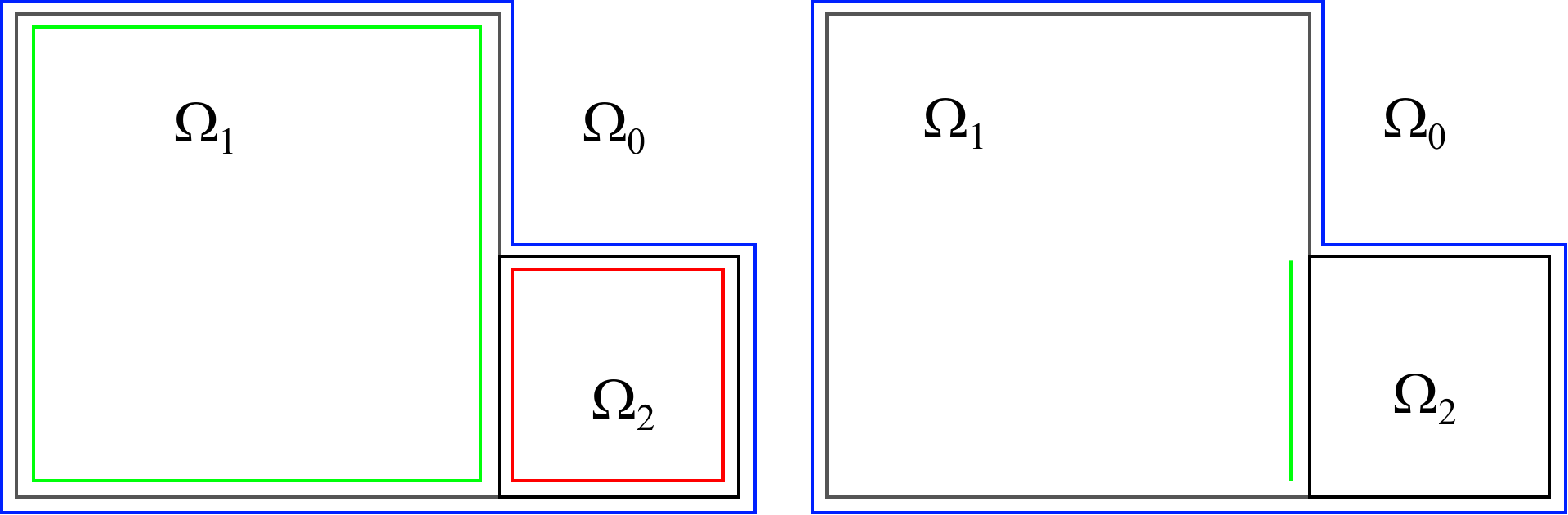}
    \caption{Example of a single-sided representation of the single-trace subspace of the multi-trace space. It is required that each edge in the skeleton mesh $\mathcal{T}_h$ is an interior edge of exactly one of the reduced boundaries $\tilde{\Gamma}_{i}$. It is possible that some of the $\tilde{\Gamma}_i$ are empty.}
    \label{fig:geo_reduction}
\end{figure}

There are many ways to construct a basis for the single trace subspace $\mathbf{H}(\dvg, \Gamma) \cap \mathbb{RT}(\mathcal{T}_{h,i})$ starting from a basis of $\bbRT(\mathcal{T}_h)$. For reasons that will become apparent later, the construction pursued here will start from a so-called \emph{single-sided} reduction of the geometry. The construction starts by identifying subsets $\tilde{\Gamma}_i \subseteq \partial \Omega_i$ that each are the union of a number of interfaces $\Gamma_{ij} := \partial \Omega_i \cap \partial \Omega_j$ and such that each edge $E \in \mathcal{E}_h$ is in the interior of exactly one of the $\tilde{\Gamma}_i$. This reduction is schematically represented in Fig.~\ref{fig:geo_reduction}. Note that is possible that some of the $\tilde{\Gamma}_i$ are empty.

The reduced boundaries $\tilde{\Gamma}_i$ inherit meshes $\tilde{\mathcal{T}}_{h,i}$ from $\mathcal{T}_{h,i}$ by restriction. Similarly, ordered sets of edges $\tilde{\mathcal{E}}_{h,i}$ and basis functions $(f^i_e)_{e \in \tilde{\mathcal{E}}_{h,i}}$ result by retaining only those $e \in \mathcal{E}_{h,i}$ that are in the interior of $\tilde{\Gamma}_i$.

Because each global edge $E \in \mathcal{E}_h$ appears in exactly one $\tilde{\mathcal{E}}_{h,i}$, iteration over the latter allows for the construction of all the degrees of freedom in the discrete single trace space.

To link the ordering and orientation of geometrifcally coinciding edges as encountered in different but adjacent domains, the sparse matrix $\mat{R}^{{f}}_{\ \tilde{f}}$ is introduced with entries:
\begin{equation}
    R^{ij}_{ee'} = \left\{ \begin{array}{ll}
        0, & \mbox{if $e$ and $e'$ are disjoint} \\
        +1, & \mbox{if $e$ and $e'$ have equal orientation} \\
        -1, & \mbox{if $e$ and $e'$ have opposite orientation} \\
    \end{array} \right.
\end{equation}
with $e \in \tilde{\mathcal{E}}_{h,i}$, $e' \in \mathcal{E}_{h,j}$, and $i,j=0,...,N$. Note that if $\bar{\Omega}_i \cap \bar{\Omega}_j = \{\}$, the corresponding block $R^{ij}$ contains only zeroes.
%
%
Now, for each $i=0,...,N$ and $e \in \tilde{\mathcal{E}}_{h,i}$ construct the functions
\begin{equation}
F^i_e := \left( \sum_{e' \in \mathcal{E}_{h,j}} R^{ij}_{ee'} f^j_{e'} \right)_{j=0}^N \in \prod_{j=0}^N \RT(\mathcal{T}_{h,i})
\end{equation}
Let $R$ be both the map that corresponds to this mapping between bases (i.e. $F^i_e = R f^i_e$), and its extension that acts on both the electric and magnetic component, i.e. $R(m,j) = (Rm, Rj)$. Likewise the notation $\mat{R}^{{f}}_{\ \tilde{f}}$ will be used for the matrix representation of both these maps. It can be seen that $R$ maps single-sided functions in $\mathbf{RT}(\tilde{\mathcal{T}}_h)$ to their unique single-trace (odd) extension in $\mathbf{RT}(\mathcal{T}_h)$. In fact, with
\begin{align}
    \RT(\mathcal{T}_h) := \spn \{ F^i_e, e \in \tilde{\mathcal{E}}_{h,i}, i \in \{0,...,N\} \}
\end{align}
we have that $\mathbf{RT}(\mathcal{T}_h) := \RT(\mathcal{T}_h) \times \RT(\mathcal{T}_h) = \mathbf{H}(\dvg, \Gamma) \cap \bbRT(\Gamma)$ is the finite element space for the single-trace Cauchy data we were looking for. 

%
The discretised single trace PMCHWT now reads: Find $(m,j) \in \mathbf{RT}(\mathcal{T}_h)$ such that, for all $(m,j) \in \mathbf{RT}(\mathcal{T}_h)$, equation \eqref{eq:ext_var} holds, or in terms of matrices:
\begin{equation}
    \mat{R}_{\tilde{f}}^{\ \ f} \mat{A}_{ff} \mat{R}^f_{\ \tilde{f}} \mat{u}^{\tilde{f}} =  \mat{R}_{\tilde{f}}^{\ \ f} \mat{e}_{f}
\end{equation}
with $\mat{R}_{\tilde{f}}^{\ \ f} = \left( \mat{R}^f_{\ \tilde{f}} \right)^{T}$, $\mat{A}_{ff}$ the matrix representation of the bilinear form $\left< \cdot , A \cdot \right>_\times$ on $\bbRT(\mathcal{T}_h) \times \bbRT(\mathcal{T}_h)$ using the standard basis, and $\mat{e}_{f}$ is the vector corresponding to the right hand side of \eqref{eq:ext_var} in the standard basis for $\bbRT(\mathcal{T}_h)$.

Using the basis for $\mathbf{RT}(\mathcal{T}_h)$ constructed above results in the classical linear system for the PMCHWT. It is well-known that the condition number of the system matrix grows as $h^{-2}$, when $h$ tends to zero. The number of iterations required by Krylov iterative solvers such as GMRES grows accordingly. In many situations of interest, and especially when junctions are present, this results in linear systems that cannot be solved on the available computational infrastructure.

When the system contains only a single domain ($N=1$), this problem can be alleviated by application of a Calder\'on preconditioner that is based on the same principles as the Calder\'on multiplicative preconditioner (CMP) for the electric field integral equation \cite{cools_calderon_2011}. The construction of the CMP is based on the availability of a finite element space of Buffa-Christiansen functions $\BC(\mathcal{T}_{h,i})$ dual to that of the primal space $\RT(\mathcal{T}_{h,i})$. This duality behaves as one would expect under taking direct products, leading to the construction of the space $\bbBC(\mathcal{T}_h)$ dual to $\bbRT(\mathcal{T}_h)$

When no junctions are present, the boundary meshes can simply inherit the ordering and orientation of their edges from those of the global skeleton mesh. The obvious reduced geometry to use is $\tilde{\Gamma}_0 = \partial \Omega_0$ and $\tilde{\Gamma}_1 = \{ \}$. These conventions leads to $R^{ij} = \operatorname{Id}$ and $F^0_e = (f^0_e, f^1_e)$. Because of the straightforward relation between the basis for the discrete single trace space and the basis of the finite element space on $\mathcal{T}_{h,i}$ it is feasible to construct the dual of the former. When junctions are present, this procedure can no longer be performed. An effective preconditioner has remained elusive.

\section{Quasi-local PMCHWT Equation}
\label{sec:qlpmchwt}

The difficulty in building a preconditioner for the PMCHWT equation stems from that fact that the selection of a subspace of the finite element space of the multi-trace space and the construction of the dual of a finite element space do not commute. This problem could potentially be alleviated by first regularising the set of equations \eqref{eq:ext_ops} and then selecting a subspace. This has to be done carefully, with the accuracy of the equation, the conditioning of the resulting system, and the invertibility at all real frequencies in mind.

In this section, the PMCHWT equation will be generalised by allowing the testing space to coincide only approximately with the single trace space. This will allow the testing space to be characterised as the image of a regularising operator, resulting in well-conditioned system matrices.

In the classic PMCHWT equation, both the trial space and testing space are given by the discrete single-trace space $\mathbf{RT}(\mathcal{T}_{h})$.
%
%
%
\begin{figure}
    \centering
    \includegraphics[width=1\linewidth]{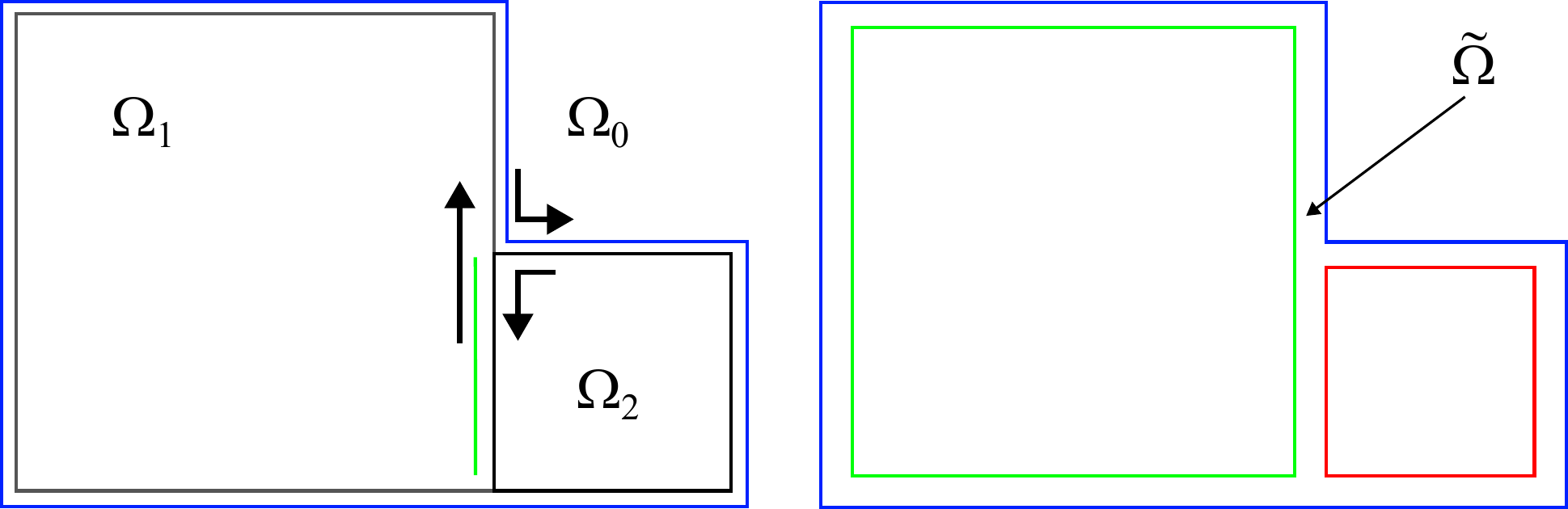}
    \caption{Left: A single-trace basis function $F^i_e$ is attached to each internal edge of the reduced geometry. This results in \emph{odd} functions that all fulfil the continuity condition. Right: the regulariser $S$ in the quasi-local PMCHWT can be interpreted as a single layer interaction in the gap domain $\tilde{\Omega}$.}
    \label{fig:delta_gap}
\end{figure}
To generalise this approach, consider the map $S$ defined by the variational formulation: Given $(m,j) \in \mathbb{H}^{-1/2}(\dvg,\Gamma)$ find $(r,s) \in \mathbb{H}^{-1/2}(\crl,\Gamma)$, such that for all $(p,q) \in \mathbb{H}^{-1/2}(\crl,\Gamma)$ it holds that
%
%
\begin{multline}
   \left<(p,q), (r,s) \right>_\times = s((p,q), (m,j) ) \\
   := \sum_{i,j=0}^N \left( s_{ij}( p_i, j_j ) - s_{ij} (q_i, m_j ) \right).
\end{multline}
%
%
where, for $s_{ij}(q_i,m_j)$ is defined by
\begin{align}
    s_{ij}(q_i,m_j) =& \delta^{-1} \iint_{\Gamma_i \times \Gamma_j} \frac{e^{-|x-y|/\delta^2}}{4\pi |x-y|} q_i(x) \cdot m_j(y) dx dy + \notag \\
    &\delta \iint_{\Gamma_i \times \Gamma_j} \frac{e^{-|x-y|/\delta^2}}{4\pi |x-y|} \dvg q_i(x) \dvg m_j(y) dx dy
\end{align}
These surface integral operators are reminiscent of -- but are not exactly equal to -- the single layer operator evaluated at the imaginary wavenumber $\kappa = \iota / \delta$. The Gaussian behaviour in the numerators is chosen to have the amplitude of these interactions drop off over a short distance, resulting in highly sparse matrices. This property of $S$ is why the single-trace method introduced here is qualified as \emph{quasi-local}. These differences with the standard single layer boundary integral operator do not jeopardise the important properties of operator $S$. The choice of the range parameter $\delta > 0$ will be further discussed in the numerical results section. An important difference between $A$ and $S$ is that $S$ is not block-diagonal with respect to the partitioning in domains.
%
%

To understand the crucial property of this map, we refer to the infinitesimal gap idea underlying the global multi-trace method. Consider the geometric setup from Fig.~\ref{fig:delta_gap}. The action of $S_{ij}$ on $m_j$ is evaluated by computing the single layer potential radiated by $m_i$ in the gap region $\tilde{\Omega}$, and taking the trace of this field on $\partial \Omega_j$. Because the gap is infinitesimally small, the field along a line connecting two opposing points in the gap is essentially constant. The opposite pointing normals of the domains meeting at any given interface $\Gamma_{ij}$ introduce a sign difference in the tangential trace. In conclusion $S$ maps from the multi-trace space $\mathbb{H}(\dvg,\Gamma)$ into the single trace space $H(\dvg,\Gamma)$. A similar intuitive argument leads to the conclusion that $S$ maps the single trace space onto $0$. For a rigorous treatment in the case of the Helmholtz equation, see \cite[Corollary 5.1]{claeys_quasi-local_2015}. This means that on the continuous level, the standard PMCHWT is equivalent to: Find $u:=(m,j) \in \mathbf{H}^{-1/2}(\dvg, \Gamma)$, such that for all $v:=(p,q) \in \mathbb{H}^{-1/2}(\dvg, \Gamma)$ it holds that
\begin{equation}
    \left< S v, \left( -\frac{1}{2} - A \right) u \right>_\times = \left< S v, u^{inc}) \right>_\times.
\end{equation}

Denote by $S_h: \bbBC(\mathcal{T}_h) \rightarrow \bbRT(\mathcal{T}_h)$ the discretisation of $S$, i.e. $\bbRT(\mathcal{T}_h) \ni (k,l) = S_h (m,j)$ if for all $(p,q) \in \bbBC(\mathcal{T}_h)$ it holds that $\left<(p,q), (k,l) \right>_\times = s((p,q), (m,j))$.
Using the bases for $\bbRT(\mathcal{T}_h)$ and $\bbBC(\mathcal{T}_h)$ introduced above leads to
\begin{equation}
    \vct{k}^f = \mat{I}^{fg} \mat{S}_{gg} \vct{m}^{g},
\end{equation}

with $\mat{I}^{fg}$ the inverse of $\mat{I}_{gf}$, where the latter is the matrix representation of the duality pairing on $\bbBC(\mathcal{T}_h) \times \bbRT(\mathcal{T}_h)$ in the standard basis. Matrix $\mat{S}_{gg}$ is the representation of the bilinear form $s( \cdot, \cdot )$ on $\bbBC(\mathcal{T}_h) \times \bbRT(\mathcal{T}_h)$ in the standard basis (often called the mixed Gram matrix). Symbols $f$ and $g$ help to track where primal and dual elements are used, respectively. Unfortunately, when junctions are present, it is no longer true that $S_h$ produces functions that are exactly in the single trace finite element space $\mathbf{RT}(\mathcal{T}_h)$.
Since it does not hold that $S_h \left( \bbBC(\mathcal{T}_h) \right) \subseteq \mathbf{H}^{-1/2}(\dvg, \Gamma)$, the self-polarity property of the single-trace space can no longer be called upon and the identity term cannot be left out.
%
%
This does not necessarily mean the image of $S_h$ cannot be used as a valid space of testing functions however. The above leads us to propose the following single trace formulation for the transmission problem: Find $u:=(m,j) \in \mathbf{RT}(\mathcal{T}_h)$, such that, for all $v:=(p,q) \in V$ it holds that
\begin{equation}
    \left< S_h v, \left( -\frac{1}{2} - A \right) u \right>_\times = \left< S_h v, u^{inc} \right>_\times.
\end{equation}
where $V$ is a suitable approximate preimage of $\mathbf{RT}(\mathcal{T}_h)$ under $S_h$ in $\bbBC(\mathcal{T}_h)$ with the same dimension as $\mathbf{RT}(\mathcal{T}_h)$
%
%
%
%
%
Here, $V$ will be chosen based on a so-called \emph{single-sided parametrisation} of the single-trace space. The starting point for this is the reduced single-sided geometry $(\tilde{\Gamma}_i)_{i=0}^N$ introduced in section~\ref{sec:classic_PMCHWT}.


The reduced geometry gives rise to boundary element spaces $\RT(\tilde{\mathcal{T}}_{h,i})$ and $\BC(\tilde{\mathcal{T}}_{h,i})$ with bases chosen to reflect a fixed ordering and orientation of the edges $\tilde{\mathcal{E}}_{h,i}$. From these spaces the direct product spaces $\mathbf{RT}(\tilde{\mathcal{T}}_h) := \prod_{i=0}^N \mathbf{RT}(\tilde{\mathcal{T}}_{h,i})$ and $\mathbf{BC}(\tilde{\mathcal{T}}_h) := \prod_{i=0}^N \mathbf{BC}(\tilde{\mathcal{T}}_{h,i})$ can be constructed, with $\mathbf{RT}(\tilde{\mathcal{T}}_{h,i}) = \RT(\tilde{\mathcal{T}}_{h,i}) \times \RT(\tilde{\mathcal{T}}_{h,i})$ and $\mathbf{BC}(\tilde{\mathcal{T}}_{h,i}) = \BC(\tilde{\mathcal{T}}_{h,i}) \times \BC(\tilde{\mathcal{T}}_{h,i})$.
%
%
In general the $\tilde{\Gamma}_i$ may have a non-void boundary. In this case, the special boundary BC functions as described in \cite{buffa_dual_2007} need to be constructed, implying that $\mathbf{BC}(\tilde{\mathcal{T}}_h)$ is not a subset of $\bbBC({\mathcal{T}}_h)$. Hence, the corresponding discretisation of $S$, denoted by $\tilde{S}_h$ and mapping $\mathbf{BC}(\tilde{\mathcal{T}}_h)$ into $\bbRT({\mathcal{T}}_h)$, is not a restriction of $S_h$. Nevertheless, the image of $\tilde{S}_h$ provides a candidate for the space of approximate single-trace functions.



The \emph{quasi-local PMCHWT} equation or QL-PMCHWT for short is defined to correspond with the choice $V = I_h \mathbf{RT}(\tilde{\mathcal{T}}_h)'$, with $\mathbf{RT}(\tilde{\mathcal{T}}_h)'$ the dual space to $\mathbf{RT}(\tilde{\mathcal{T}}_h)$ and $I_h$ the map from $\mathbf{RT}(\tilde{\mathcal{T}}_h)'$ to $\mathbf{BC}(\tilde{\mathcal{T}}_h)$ defined by $v := I_h \nu$ if for all $w \in \mathbf{BC}(\tilde{\mathcal{T}}_h)$ it holds that $\left< w, v \right>_\times = \nu(w)$. In other words, find $u:=(m,j) \in \mathbf{RT}(\mathcal{T}_h)$, such that, for all $\nu \in \mathbf{RT}(\tilde{\mathcal{T}}_h)'$ it holds that
\begin{equation}
    \left< \tilde{S}_h I_h \nu, \left( -\frac{1}{2} - A \right) u \right>_\times = \left< \tilde{S}_h I_h \nu, u^{inc} \right>_\times.
\end{equation}
%
%
Using the basis $\left( f \rightarrow \left< f^i_e, f \right>_\times \right)_{e \in \tilde{\mathcal{E}}_{h,i}}$ for $\RT(\tilde{\mathcal{T}}_{h,i})'$ leads to the matrix equation: Find $\mat{w}^{\tilde{f}} \in \mathbb{C}^{2| \tilde{\mathcal{E}}_h} |$ that solves
\begin{equation}
    \left( \mat{I}^{fg} \mat{S}_{g\tilde{g}} \mat{I}^{\tilde{g}\tilde{f}} \right)^{T} \left( -\frac{1}{2} \mat{I}_{ff} - \mat{A}_{ff} \right) \mat{R}^f_{\ \tilde{f}} \vct{w}^{\tilde{f}} = \left( \mat{I}^{fg} \mat{S}_{gg} \mat{I}^{\tilde{g}\tilde{f}} \right)^{T} \vct{e}_f
\end{equation}
or, equivalently,
\begin{equation}
    \mat{I}^{\tilde{f}\tilde{g}} \mat{S}_{\tilde{g}g} \mat{I}^{gf} \left( -\frac{1}{2} \mat{I}_{ff} - \mat{A}_{ff} \right) \mat{R}^f_{\ \tilde{f}} \vct{w}^{\tilde{f}} = \mat{I}^{\tilde{f}\tilde{g}} \mat{S}_{gg} \mat{I}^{gf} \vct{e}_f
\end{equation}
Matrix $\mat{S}_{\tilde{g}g}$ is the representation of the bilinear form $s$ on $\mathbf{BC}(\tilde{\mathcal{T}}_h) \times \mathbb{BC}(\mathcal{T}_h)$ in the standard basis, and has size $2|\tilde{\mathcal{E}}_h| \times 2|\mathcal{E}_h|$ instead of the roughly twice larger $2|\mathcal{E}_h| \times 2|\mathcal{E}_h|$, and is essentially sparse.

The above expression reveals why one would be willing to part from the exact single trace space as a testing space. The adjoint of the map $\tilde{S}_h$ used in the parametrisation of the testing space can double as regulariser for the Calder\'on projector at the heart of the formulation!

It is known that for the exact PMCHWT in the absense of junctions, $\mat{I}^{\tilde{f}\tilde{g}} \mat{S}^{T}_{\tilde{g}g} \mat{I}^{gf}$ is an effective preconditioner \cite{cools_calderon_2011}. This can be expected to remain true if the residual contribution of the identity term remains sufficiently small as $h \rightarrow 0$. Because the regulariser is non-square and because the coercivity of $A$ is not sufficient to argue the $\inf \sup$-stability of the factor in parentheses, the standard functional theory from \cite{hiptmair_operator_2006} does not apply. In this contribution, numerical experiments will be presented to assess the effectiveness of the regularisation and the stability of the discretisation.


Another point that merits numerical investigation is whether the scheme is free from internal resonances. This is also linked to the behaviour of the residual contribution of the identity terms.

There may be some concern as to whether the method introduced here is truly deserving of the moniker \emph{single trace}. We provide some arguments in its defence:
\begin{itemize}
    \item The solution $\mat{w}^{\tilde{f}}$, or rather $\mat{R}^f_{\ \tilde{f}} \vct{w}^{\tilde{f}}$, lies exactly in the single trace subspace. This property is not fulfilled by the solution to the discrete multi-trace PMCHWT equation \cite{lasisi_fast_2022} or the discrete multi-trace single source integral equation \cite{munger_resonant_2024}.
    \item For configurations without junctions such as single domain problems, problems featuring multiple free-floating domains, or problems featuring an \emph{onion-shaped} geometry, the quasi-local PMCHWT reduces exactly to the Calder\'on preconditioned classic - single trace - PMCHWT.
    \item The operator $S$ acts on functions supported by the surface $\partial \Omega_0$, which is the true boundary of $\Omega_0$ and much smaller than $\partial \tilde{\Omega}$. In fact, the domain $\Omega_0$ does not play a special role in this method, unlike in global multi-trace methods. Interactions in $\tilde{\Omega}$ on the other hand, only travel over a distance in the order of $\delta$. No compression algorithms such as the fast multipole method \cite{chew_fast_2001} or an H-matrix method \cite{borm_introduction_2003} are needed. In addition, the regulariser is real-valued, further decreasing computation time and storage costs. These advantages counteract to a large extent the increase in computational costs due to the introduction of the barycentric refinement.
    \item The QL-PMCHWT features an identity term that is absent in the classic PMCHWT. Even though this is superficially reminiscent to the identity term that appears in the global multi-trace method in positions in the system that correspond with inter-domain interactions, it is of a different nature. The identity that appears in the QL-PMCHWT is a \emph{true} identity operator and not a \emph{geometric} identity \cite{lasisi_fast_2022} that produces non-zero off-diagonal contributions. This means that whereas the construction of the multi-trace matrices requires bespoke modifications to the assembly loop, the QL-PMCHWT matrices can be built by most existing boundary element method implementations without changes to the code.
\end{itemize}

\section{Numerical Results}

Throughout the numerical results section, the excitation is chosen to be a plane wave with signature $e^{inc}(x) = (1,0,0)^{T} \exp \left( -\iota \kappa_0 x_3 \right)$, where $\kappa_0$ is the wavenumber in the background medium. All results have been produced by Julia package BEAST.jl \cite{cools_krcoolsbeastjl_2025}.

\subsection{Comparison with Mie series}

\begin{figure}
    \centering
    \includegraphics[width=0.9\linewidth]{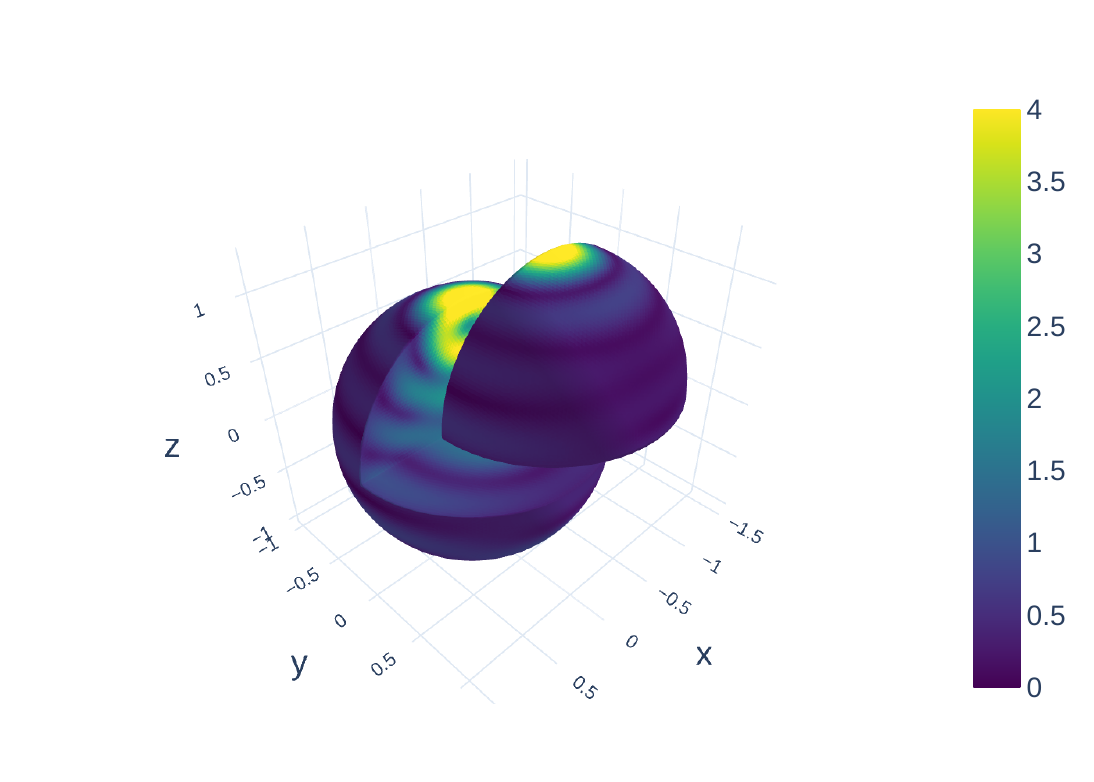}
    \caption{Sphere partitioned in a domain covering three quadrants $\Omega_1$ and a domain covering the remaining quadrant $\Omega_2$. All interior domains are occupied by the same material to allow comparison with the Mie series. Domain $\Omega_2$ is shifted out for clarity. Colours refer to the norm of the tangential electric field [Volt/meter]. The reduced geometry is chosen as $(\partial \Omega_0, \partial \Omega_1 \cap \partial \Omega_2, \{\})$}
    \label{fig:p6_facecurs}
\end{figure}

\begin{figure}
    \centering
    \includegraphics[width=1\linewidth]{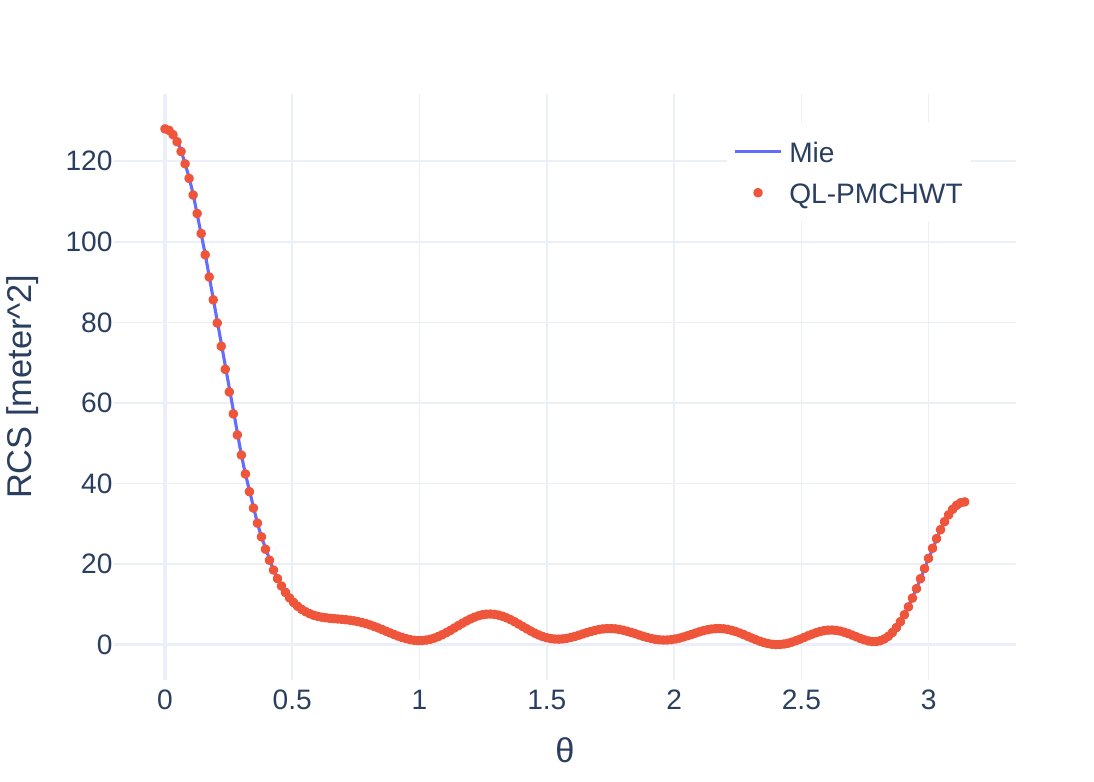}
    \caption{The radar cross section for the sphere shown in Fig.~\ref{fig:p6_facecurs} at $\kappa_0 = 6$ meter computed by the Mie series and by the quasi-local PMCHWT agree well.}
    \label{fig:p6_farfield}
\end{figure}

To verify correctness of the algorithm on a non-trivial example, a uniform sphere characterised by permittivity and permability $(\epsilon, \mu) = (3 \epsilon_0, \mu_0)$ is artificially split in two domains, as in Fig.~\ref{fig:p6_facecurs}. The wavenumber in the background medium is set to $\kappa_0 = 6.0$ per meter. The radar cross section computed by solving the transmission problem with the quasi-local PMCHWT method and that obtained by evaluation of the Mie series \cite{hofmann_sphericalscattering_2025} agree well (Fig.~\ref{fig:p6_farfield}). The quasi local PMCHWT was solved for a mesh size of $0.028$ meter, resulting in $147,030$ degrees of freedom. GMRES converged in $495$ iterations within a relative residual error of $2e-5$.

\subsection{Rate of Convergence}
\label{subsec:rateofconvergence}

\begin{figure}
    \centering
    \includegraphics[width=0.9\linewidth]{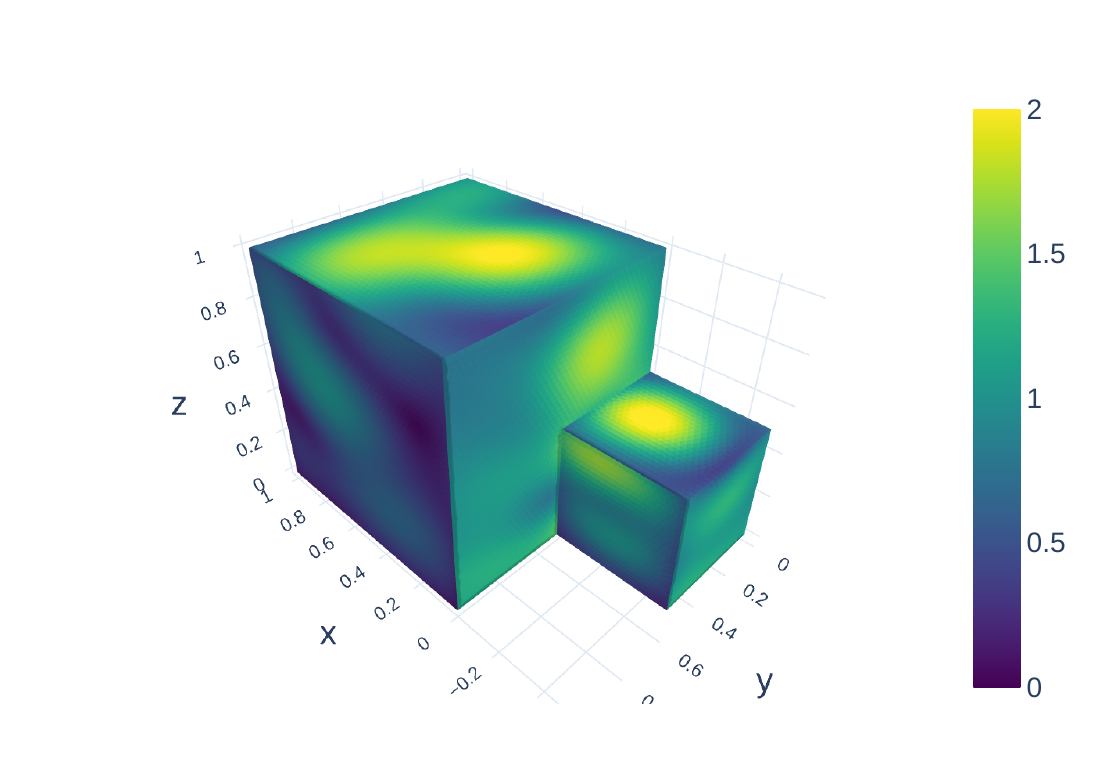}
    \caption{Geometry comprising two adjacent boxes of different size. The colour scale refers to the norm of the normalised magnetic trace $\eta_0 n \times h$ [Volt/meter] at mesh size $0.24$ meter for the experiment described in section \ref{subsec:rateofconvergence}.}
    \label{fig:p5_facecurs}
\end{figure}

Consider the geometry comprising two juxtaposed cubes (Fig.~\ref{fig:p5_facecurs}): a larger cube occupying a domain $\Omega_1$ with side length $1$ meter and $(\epsilon_1, \mu_1) = (2 \epsilon_0, 1 \mu_0)$, and a smaller cube $\Omega_2$ occupying a domain $\Omega_2$ with side length $0.5$ meter and $(\epsilon_2, \mu_2) = (4 \epsilon_0, 1 \mu_0)$. The single-sided reduction is chosen to be $\tilde{\Gamma}_0 = \Gamma_0$, $\tilde{\Gamma}_1 = \Gamma_1 \cap \Gamma_2$, and $\tilde{\Gamma}_2 = \{\}$. The transmission problem is solved at $\kappa_0=6$ per meter with the quasi-local PMCHWT for mesh sizes decreasing from $h=0.25$ meter to $h=0.024$ meter. The solution at the smallest value for $h$ is used as a reference solution. For all other values for $h$, the energy norm of the difference between the solution and the reference solution is computed. To compute the energy norm, consider the symmetric bilinear form
\begin{equation}
    \hat{t}((p,q),(m,j)) := \sum_{i=0}^N \left( \hat{t}_{i}(p_i, m_i) + \hat{t}(q_i, j_i) \right)
\end{equation}
with
\begin{align}
\hat{t}_{ij}(p_i, m_i) & = \kappa_0 \iint_{\Gamma_i^2} \frac{e^{-\kappa_0 |x-y|}}{4\pi |x-y|} p_i(x) \cdot m_i(y) dxdy + \notag \\
& = \frac{1}{\kappa_0} \iint_{\Gamma_i^2} \frac{e^{-\kappa_0 |x-y|}}{4\pi |x-y|} \dvg p_i(x) \dvg_i(y) dxdy \notag.
\end{align}
and its matrix representation $\hat{\mat{T}}_{ff}$. The energy norm of a vector $\mat{u}^{f}$ of expansion coefficients is $\sqrt{\mat{u}^{f*} \hat{\mat{T}}_{ff} \mat{u}^{f}}$, where the asterisk denotes taking the complex adjoint.

\begin{figure}
    \centering
    \includegraphics[width=1.0\linewidth]{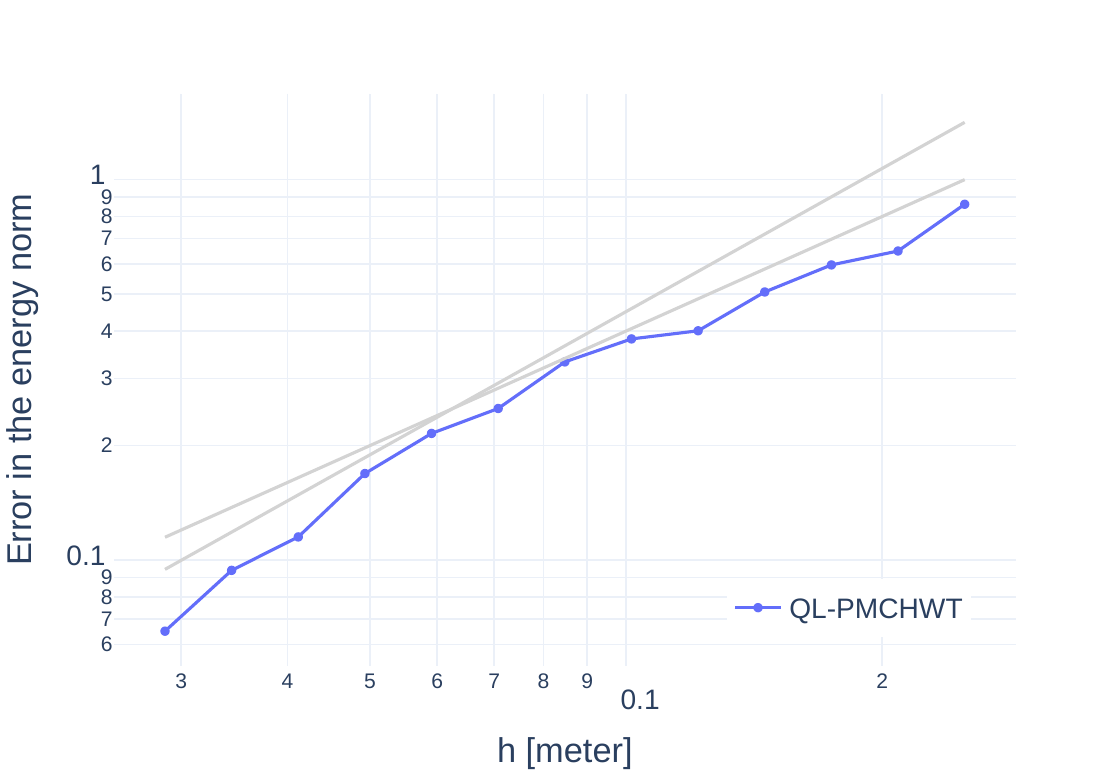}
    \caption{The energy norm of the difference between the QL-PMCHWT solution of the transmission for the geometry shown in Fig.~\ref{fig:p5_facecurs} at $h$ and $h_{ref}=0.02$ meter. For comparison, curves proportional to $h^{1}$ and $h^{5/4}$ are shown.}
    \label{fig:p5_energy_error}
\end{figure}

This procedure reveals an asymptotic rate of convergence of about $h^{1.25}$ (Fig.~\ref{fig:p5_energy_error}). The classic PMCHWT method did not converge for mesh size $h_{ref}$ so its solution could not be used as a reference. The downside of using the error criterion presented here, is that it relies on a reference solution computed by the method-under-test that is assumed trustworthy. This shortcoming will be addressed in the next section.

\subsection{Extinction Error}

\begin{figure}
    \centering
    \includegraphics[width=1\linewidth]{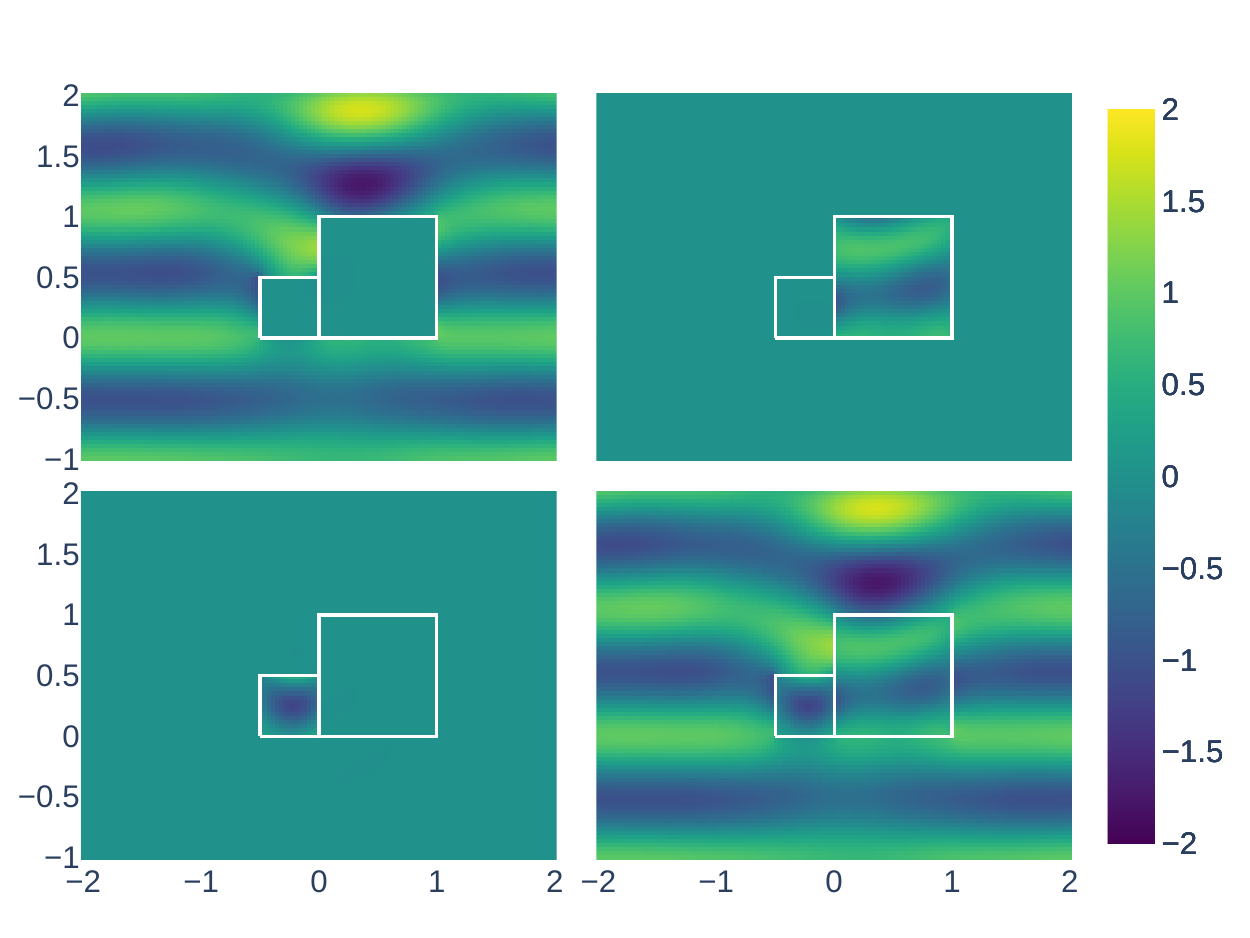}
    \caption{Real part of the $x_2$-component of the electric field [Volt/meter] reconstructed from the Cauchy data on $\partial \Omega_i$ using the corresponding material on a grid at $x_2=0.25$ meter that contains points both in and outside $\Omega_i$. The traces $(m_i,j_i)$ fulfil the extinction criterion and so must belong to a Maxwell solution in $\Omega_i$. The bottom right subplot shows to total field, which is exactly continuous at the interfaces because it originates from a single trace function.}
    \label{fig:p5_nearfield}
\end{figure}

\begin{figure}
    \centering
    \includegraphics[width=1\linewidth]{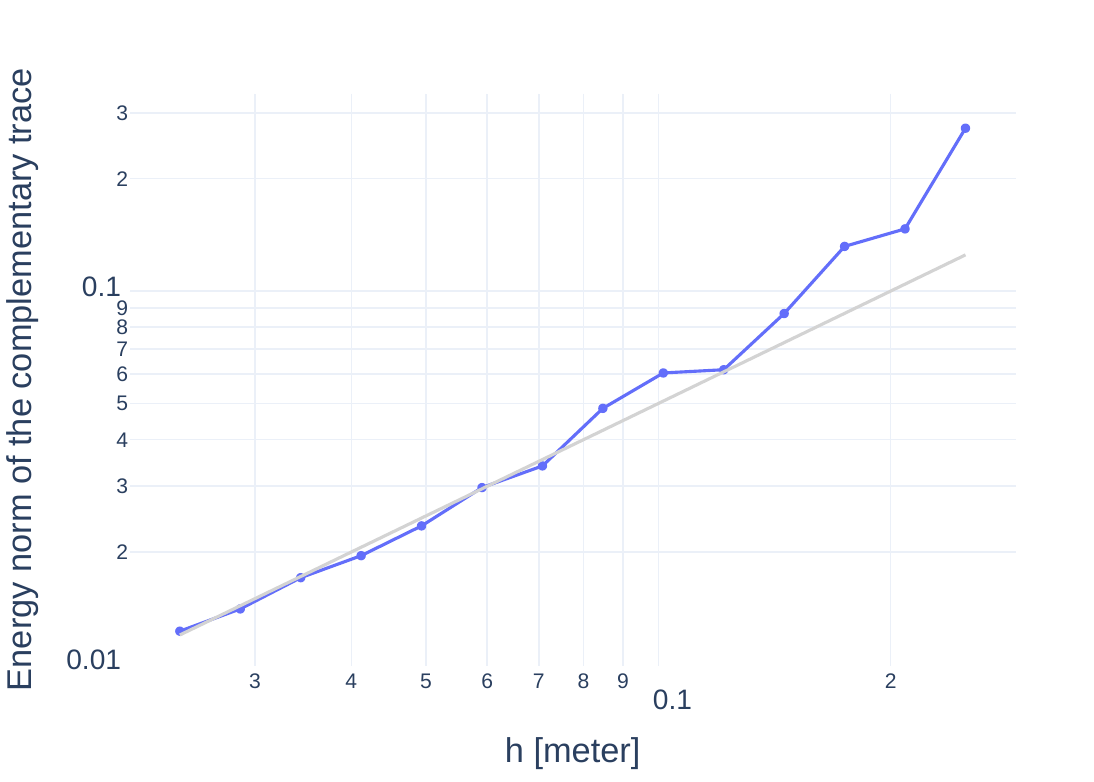}
    \caption{Energy norm of the complementary trace of the approximate solution at mesh size $h$, which for the exact solution should be zero. Computation of this measure involves projection on the dual finite element space. For comparison, the line $h^1$ is shown as well.}
    \label{fig:p5_extinction_error}
\end{figure}

As an alternative verification for the correctness and convergence rate of the algorithm, the extinction property is verified for the simulations done in section~\ref{subsec:rateofconvergence}. For the simulation at $h=0.024$ meter, the Stratton-Chu formula in $\Omega_i$, $i=0,1,2$ is used on a grid containing points both in and outside of the domain of interest. For points in $\Omega_i$ the formula leads to the reconstruction of a Maxwell solution with respect to the corresponding material properties. For points in the complement $\mathbb{R}^3 \setminus \bar{\Omega}_i$ contributions from the single and double layer potential cancel.

A more quantitative study is achieved by testing the complementary trace on the $\partial \Omega_i$, i.e. computing the quantity
\begin{equation}
    \mat{v}_{f} := \left( -\mat{A}_{ff} - \frac{1}{2} \mat{I}_{ff} \right) \mat{R}^f_{\ \tilde{f}} \vct{w}^{\tilde{f}} - \mat{e}_{f}
\end{equation}
Note that $\mat{w}^{\tilde{f}}$ being the solution to the QL-PMCHWT does not imply that $\mat{v}_{f}$ is zero. The projection on the dual basis using the inverse duality matrix between primal RWG functions and dual BC functions yields an approximation of the complementary trace as an expansion in BC functions: $\mat{v}^{g} := \mat{I}^{gf} \mat{v}_{f}$. Finally, the energy norm of $\mat{v}^{g}$ is computed (following the procedure in previous section but now applied to a vector of expansion coefficients w.r.t. the dual finite element space). The advantage of using this \emph{extinction error} measure is that it does not rely on a reference solution. The disadvantages are that it requires projection on the dual basis, which is known to exhibit inferior approximation strength, and that the measure is intimately related with the variational formulation of the scattering problem and its discretisation. The results shown in Fig.~\ref{fig:p5_extinction_error} suggest a convergence rate of $h^1$.

\subsection{Dense grid regime}

\begin{figure}
    \centering
    \includegraphics[width=1\linewidth]{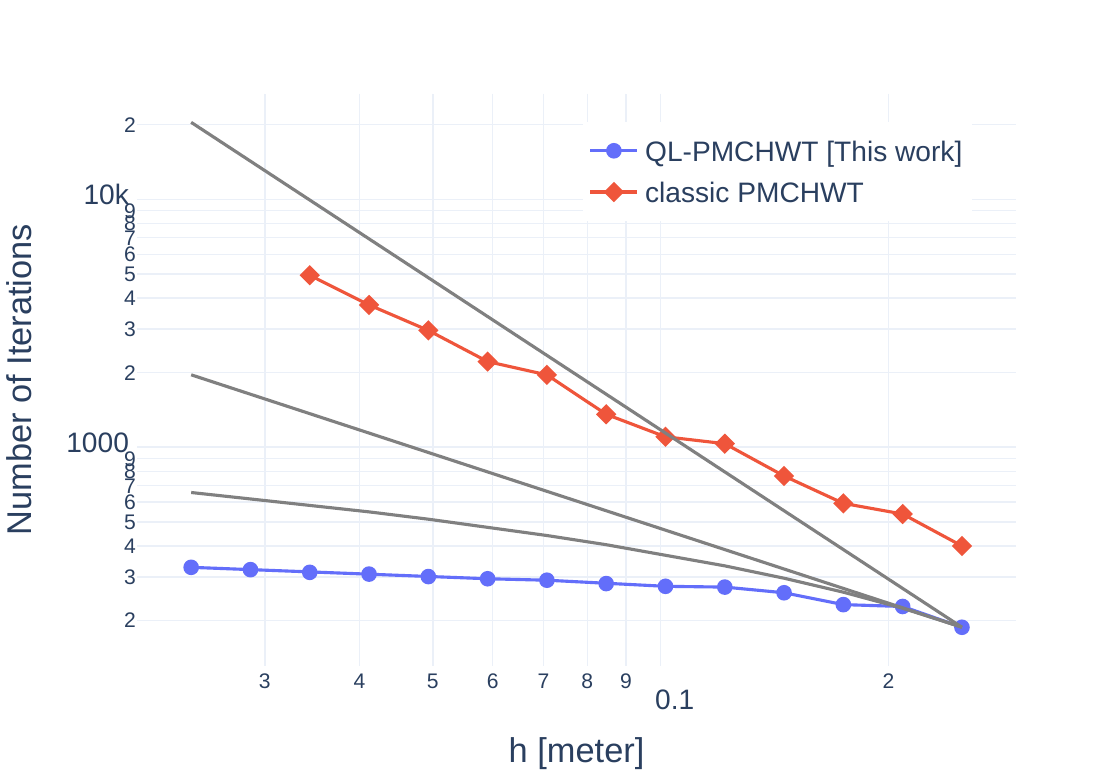}
    \caption{Numer of iterations required for GMRES convergence up to a tolerance of $2e-5$ for the classic PMCHWT and the quasi-local PMCHWT introduced here. The number of iterations for the QL-PMCHWT grows only slowly as $h$ decreases. For comparison, curves proportional to $h^{-1}$ and $h^{-2}$ are shown together with the fitted logarithm $-200 \ln h - 89$.}
    \label{fig:p5_iterations}
\end{figure}

For the same sequence of simulations as in section~\ref{subsec:rateofconvergence}, the number of iterations for GMRES to converge to a relative precision of $2e-5$ was recorded for the classic PMCHWT and the quasi-local PMCHWT. The results are shown in Fig.~\ref{fig:p5_iterations}, together with the asymptotics $h^{-1}$, $h^{-2}$, and the curve $-200 \ln h - 89$. The number of iterations for the classic PMCHWT grows at least as fast as $h^{-1}$. For the QL-PMCHWT the number of iterations grows only slowly, increasing from $188$ to $328$ as $h$ drops a full decade from $0.25$ meter to $0.024$ meter. Even though the number of iterations is not asymptotically constant, it increases only very slowly.

\subsection{Susceptibility to spurious resonances}

\begin{figure}
    \centering
    \includegraphics[width=1\linewidth]{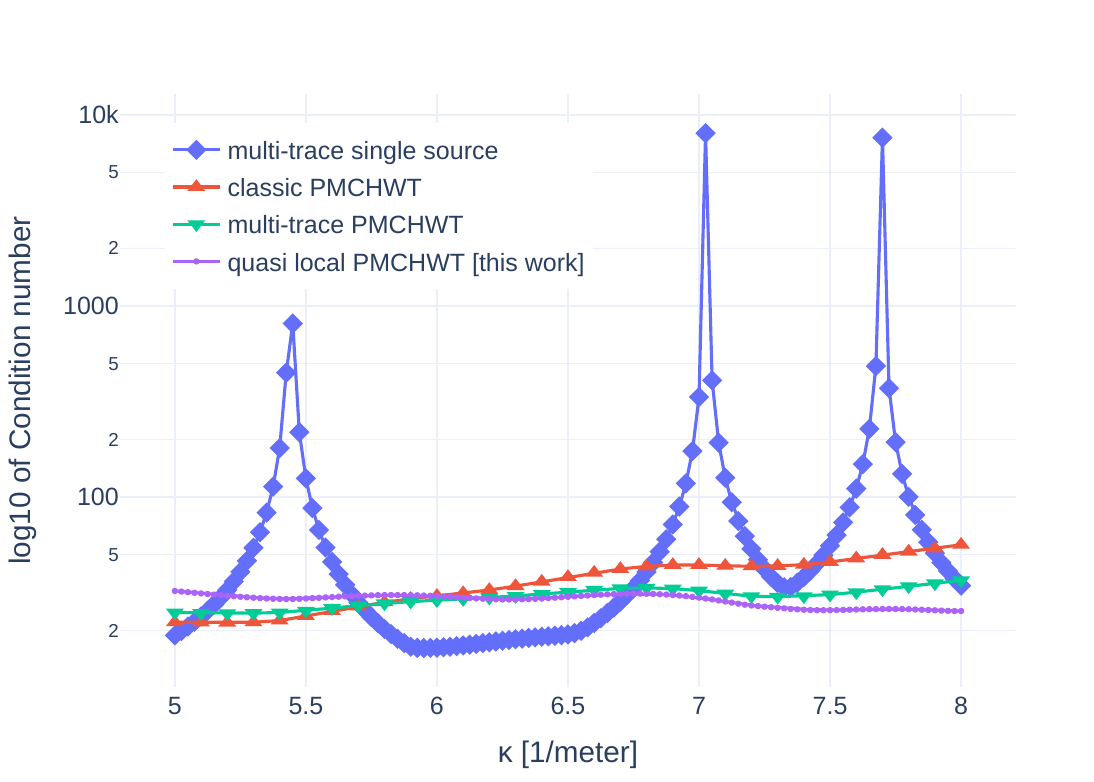}
    \caption{System matrix condition number versus the wavenumber in $\Omega_0$ for the multi-trace single-source equations, the  (local) single-trace PMCHWT equations, the multi-trace PMCHWT equation, and the quasi-local PMCHWT equation. The well-documented susceptibility of the MT-SS equation corresponds to sharp spikes in the condition number. The quasi-local PMCHWT [this work] has a bounded condition number.}
    \label{fig:p3_cn_vs_kappa}
\end{figure}

For the same geometry as in section~\ref{subsec:rateofconvergence} but now occupied by materials $(\epsilon_1, \mu_1) = (\sqrt{2} \epsilon_0, \sqrt{2} \mu_0)$ and $(\epsilon_2, \mu_2) = (2 \epsilon_0, 2 \mu_0)$, the tranmission problem is solved with the classic single trace PMCHWT \cite{yla-oijala_surface_2005, kolundzija_electromagnetic_1999}, the multi-trace PMCHWT \cite{lasisi_fast_2022}, the multi-trace single source integral equation \cite{munger_resonant_2024}, and quasi local single trace PMCHWT [this work]. The wavenumber is varied from $5$ per meter up to $8$ per meter. At each value of the wavenumber, the condition number of the system matrix is computed. For the classic single trace PMCHWT the matrix is \emph{preconditioned} by the inverse of the system matrix at the purely imaginary wavenumber $\iota \kappa$. This is done to remove the effects of dense grid breakdown and to make the results for the single trace PMCHWT more readily comparable with those for the other three methods. It is known that the single trace PMCHWT and multi-trace  PMCHWT remain uniquely solvable, also at wavenumbers that correspond to internal resonances. On the other hand, it is well known that the single source integral equation (without further manipulations as in \cite{valdes_calderon-preconditioned_2011,costabel_kleinmanmartin_2011,munger_resonant_2024}) becomes singular at certain wavenumbers.

The results in Fig.~\ref{fig:p3_cn_vs_kappa} demonstrates that the quasi-local PMCHWT, just like the exact PMCHWT and the multi-trace PMCHWT, is free from resonances.

\subsection{Necessity of the identity term}

\begin{figure}
    \centering
    \includegraphics[width=1\linewidth]{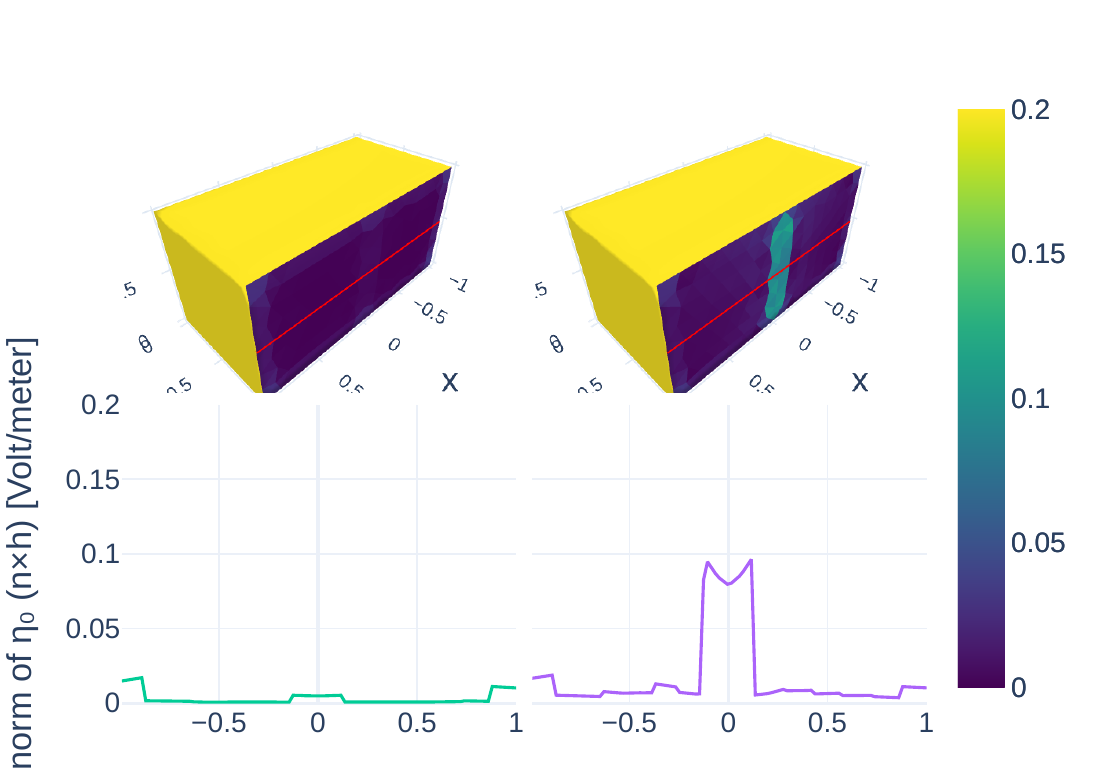}
    \caption{Norm of the normalised magnetic trace $\eta_0 (n \times h)$ [Volt/meter] for the solution of the QL-PMCHWT equation at mesh size $0.15$ meter with the identity term retained (left) and discarded (right). When the identity term is discarded, artefacts appear near the junctions present in the geometry. The colour scale has been manually limited to $[0, 0.2]$ to more clearly bring out this effect. The bottom two graphs plot the values along the red line indicated in the 3d plots.}
    \label{fig:p7_artefact}
\end{figure}

The continuous variational formulation of the quasi-local PMCHWT equation is equivalent to the local PMCHWT equation. This means that there is no difference whether the identity term in the Calder\'on projector is left in or not. Upon discretisation this changes, and two different non-equivalent systems emerge. Consider the transmission problem for two adjacent cubes, occupied by the background medium. In this special case, the solution for the traces should be constant on each face of the cuboid. Fig.~\ref{fig:p7_artefact} shows that this is not true if the identity term is left out.

\subsection{Choice of the regulariser range}

\begin{figure}
    \centering
    \includegraphics[width=1\linewidth]{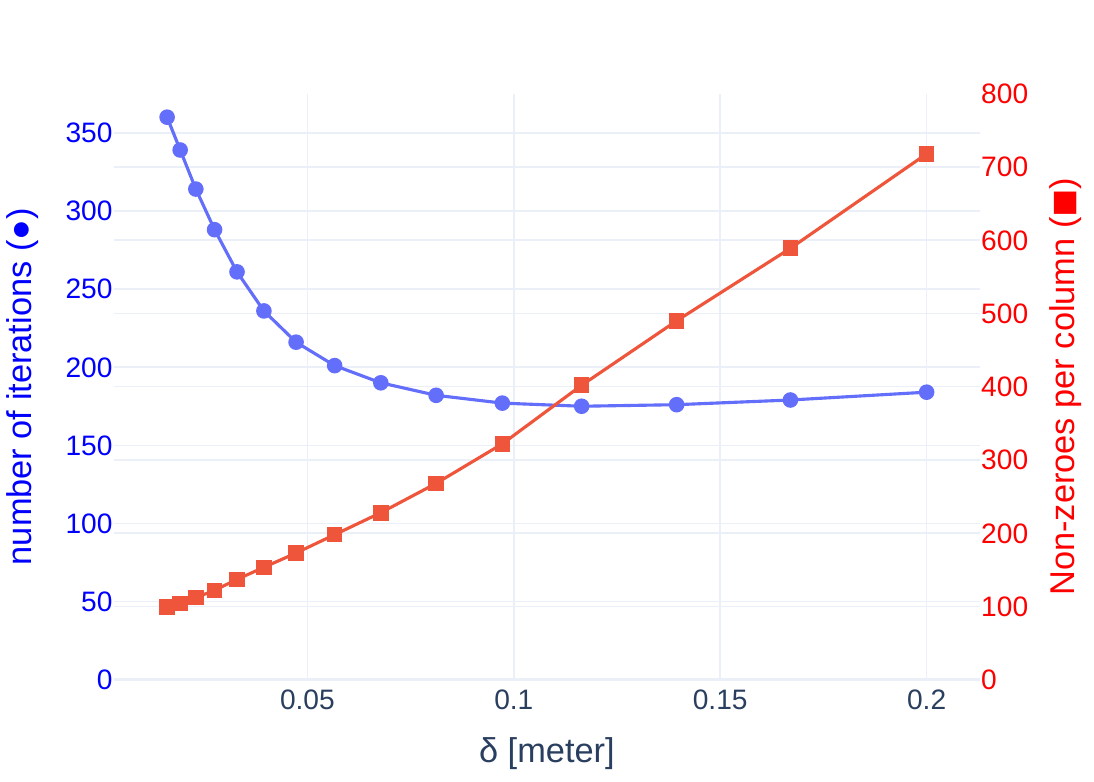}
    \caption{Number of iterations and number of non-zero entries for different values of the truncation parameter $\delta$. The value $\delta \approx h$ provdes a good trade-off between memory savings and Krylov convergence rate.}
    \label{fig:p5_numit_vs_delta}
\end{figure}

The regulariser used here exhibits the same sign signature and singularity behaviour as the single layer operator evaluated at a purely imaginary wavenumber. The nominators of the integrand, however, are modified to limit the range over which the underlying interaction acts. This leads to a sparse matrix that can be efficiently multiplied with a vector without the need to use advanced matrix compression algorithms. It also mitigates the increased computational cost of introducing the barycentric refinement of the mesh, on which the Buffa-Christiansen basis functions are defined. In this section it is investigated how small the regulariser support can be chosen without nullifying its beneficial effects on the system matrix spectrum and the resulting decrease in number of iterations.

At a fixed mesh parameter of $h=0.101$ meter, the problem from section~\ref{subsec:rateofconvergence} is solved with $\delta$ ranging decreasing from $0.2$ meter to $0.016$ meter. Interactions over distances larger than $3.5\delta$ are set to zero. For each value of $\delta$, the number of iterations required for the Krylov solver the converge up to a relative tolerance of $2e-5$ is recorded, together with the number of non-zero entries in each column of the regulariser.
The results in Fig.~\ref{fig:p5_extinction_error} suggest a value for $\delta$ between $h$ and $h/2$.

\section{Conclusion}
The quasi-local PMCHWT is introduced, a novel single trace integral equation method for the modelling of transmission through composite penetrable systems. By using a judiciously chosen parametrisation of the approximate single-trace space as testing space, the method yields both a good error convergence rate and a slowly growing number of Krylov solver iterations. The method is free from resonances and its computational cost can be optimised by leveraging a short range varaition of the single layer potential. Further research will focus on high contrast and low frequency behaviour of the approach.


\bibliographystyle{unsrt}
\bibliography{references}

\end{document}